# Study of Centrality Dependence of Transverse Momentum Spectra of Hadrons and the Freeze-out Parameters at $\sqrt{s_{NN}}$ = 62.4, 130 and 200 GeV

Saeed Uddin, Riyaz Ahmed Bhat[*] and Inam-ul Bashir

*Department of Physics, Jamia Millia Islamia (Central University)*

New Delhi-110025

## Abstract

We attempt to describe the rapidity distribution of p, p̄, $K^+$ and $\overline{K}$ for the most central Au+Au collisions at $\sqrt{s_{NN}}$ = 62.4 GeV, 130 GeV and 200 GeV. The transverse momentum spectra of strange as well as non-strange hadrons e.g. p, p̄, $K^+$, $K^-$, Λ, $\overline{\Lambda}$, Ξ, $\overline{\Xi}$ and ($\Omega^- + \Omega^+$) are studied for the whole centrality classes at all the three RHIC energies. The experimental data of the transverse momentum spectra and the rapidity distributions are well reproduced. This is done by using a statistical thermal freeze-out model which incorporates the rapidity (collision) axis as well as transverse direction boosts developed within an expanding hot and dense hadronic fluid (fireball) till the final freeze-out. We determine the thermo-chemical freeze-out conditions particularly in terms of temperature, baryon chemical potential and collective flow effect parameters for different particle species. The parameters indicate occurrence of freeze-out of the singly and doubly strange hyperon species at somewhat earlier times during the evolution of the fireball. Dependence of the freeze-out parameters on the degree of centrality is also described and it is found that the kinetic temperature increases and collective flow effect decreases with decreasing centrality at all energies studied. The nuclear transparency effect is also studied and it is clear from our model that the nuclear matter becomes more transparent at the highest RHIC energy compared to lower RHIC energies. The contribution of heavier hadronic resonance decay is taken into account.

[*]*Email: riyaztheory@gmail.com*

## 1.1 Introduction

The experiments at RHIC and LHC involving study of nuclear matter under extreme conditions of temperatures and densities offer an opportunity for insight into the properties of strongly interacting hadronic matter as described by non-perturbative quantum chromodynamics (QCD). The important part of this study is the search of a new state of matter called Quark Gluon Plasma (QGP) in which the quarks and gluons move over a special region having dimensions greater than the dimensions of a hadron. This state of matter is believed to be present in nature only in astrophysical objects like the core of neutron stars and collapsing supernovae [1]. The long range properties of nuclear matter can be studied in experimental labs only by means of heavy-ion collisions which form comparatively small systems over short times. By varying the bombarding energy as well as projectile and target combinations including the centrality of the collisions, it is possible to create systems of different energy, baryon density and sizes. The centrality (related to the impact parameter) is a key parameter in the study of the properties of QCD matter at extreme temperature and energy density because it is directly related to the initial overlap region of the colliding nuclei [2]. When the density of the baryonic matter is low, the nuclear liquid has a possibility to change into a phase of nucleons [1]. At high densities the phase transition to the quark gluon plasma (QGP) is predicted to occur, where quarks are de-confined and the chiral symmetry is restored. We define the Quark-Gluon plasma (QGP) state to be a locally thermally equilibrated state of matter consisting of de-confined quarks and gluons interacting through color charge. In the transition to de-confined QGP, the hadrons would lose the ownership of their constituent partons, which would be free to propagate across the whole extent of the medium. Indeed the concept of a hadron would no longer apply. Identifying and studying the properties of these phases is a challenging task that requires knowledge of the evolution of the hadronic phase and its macroscopic properties. Experimental phenomenology provides valuable input to theory, which at present is not able to model the complete dynamics of the evolution of the system formed in a heavy-ion collision [1]. This is mainly because of the complex and non-perturbative nature of the processes involved. Nevertheless, the high energy collisions between heavy nuclei provide us a valuable and unique opportunity to study the properties of the nuclear matter under extreme conditions where, after the initial collisions between two heavy nuclei, a fireball is formed which is very hot and dense. This consists of the incident (participating) partons in the beam and the newly produced (secondary) ones. The initial density and temperature of these partons (quarks and gluons) before the formation of hadrons (hadronization) in such systems are very high. The hadrons thus created from this hot and dense partonic matter, or the so called soup of quarks and gluons form a very hot, dense and strongly interacting matter. Hence the predicted quark – hadron phase transition occurs. As the hadrons require some finite time to be formed, the matter spends some time in the mixed phase where quarks and gluons co-exist with hadrons. The expansion is likely to be isothermal in this phase and the latent heat is absorbed in the conversion of the degrees of freedom of quarks and gluons into hadronic degrees of freedom. The constituents of this hot and dense hadronic matter (hadrons) continue to interact leading to further evolution of the system. This state of interacting hadrons, the so called hadronic resonance gas (HRG), may last for a few femto seconds ($\tau \sim 5 - 10$ fm/c). During this period various inelastic and elastic hadronic reactions take place and finally the hot matter reaches a stage of freeze-out where a gas of nearly non-interacting hadrons exists. This stage is called the thermo-chemical freeze-out where the particle number changing processes stop and the particle ratios and particle spectra are frozen in time. The particles then finally stream out towards the detectors. Within the framework of the Statistical Model [3,4], the measured ratios can be used to constrain the system temperature and the baryonic chemical potential, $\mu_B$, at chemical freeze-out. The above mentioned Statistical Models predict that the system is in thermal and chemical equilibrium at this stage.

The produced hadrons may carry information about the collision dynamics and the subsequent space-time evolution of the system till the occurrence of the final freeze-out. The corresponding freeze-out conditions, such as the temperature and the collective hydrodynamical flow effects (transverse and longitudinal expansions) for each hadronic species at the time of their freeze-out, can be obtained from the analysis of their respective transverse momentum distributions. Some hydrodynamical models [5,6] that include radial flow have successfully described the measured $p_t$ distributions in Au+Au collisions at $\sqrt{s_{NN}} = 130$ GeV [7,8]. The $p_T$ spectra of identified charged hadrons below 2GeV/c in central collisions have been well reproduced in some models by two simple parameters: transverse flow velocity $\beta_T$ and thermal freeze-out temperature T [8] under the assumption of thermalization. Some statistical models have successfully described the particle abundances at low $p_T$ [9-11]. Also in our earlier attempt [12], we have studied the rapidity and transverse momentum distributions of hadrons produced in central Au-Au collisions at Highest RHIC energy $\sqrt{s_{NN}} = 200$ GeV.

In this paper we present the study of transverse momentum distribution of hadrons produced at different RHIC energies at different centralities using the same unified thermal freeze-out model [12] which incorporates *both* longitudinal as well as transverse direction boosts. The integrated yields of the various hadronic species coming from the different parts (or elements) of the hydro-dynamically expanding hadronic matter is calculated in the mid-rapidity (|y| < 0.05) region. We present a study of the centrality dependence of thermal freeze-out temperature and collective flow velocity at the final hydrothermal freeze-out in Au+Au collisions for $\sqrt{s_{NN}} = 62.4$, 130 and 200 GeV measured by the RHIC experiments. Collective or the hydrodynamical flow effect is observed for several particles emerging from a nuclear reaction in heavy ion collisions and refers to the directed movement of a large number of particles either in a common direction.

## 1.2 Results and Discussions

In a previous analysis [13] the pion rapidity distribution was used to fix $\sigma$, the width of the distribution. The same was used for obtaining the rapidity spectra of the other hadrons (protons and antiprotons) also. A common parameterization for the local thermal temperature was used for obtaining a fit to the rapidity spectra of these particles. No attempt was made to describe the transverse momentum spectra of these hadrons. In contrast we have reproduced the transverse momentum as well as the rapidity spectra of the various particles simultaneously in the framework of our unified statistical thermal freeze-out model [7, 12]. The thermal temperature T and the flow parameter $\beta_T^0$ have a pronounced effect on the transverse momentum distribution of the hadrons but have a weak effect on the rapidity spectra. Hence these are fixed by obtaining a best fit to the *given* particle's transverse momentum spectrum obtained in the RHIC experiment. This is done for each hadron separately and at all energies and centralities. The remaining parameters i.e. $\sigma$ and $r_0$ have a very weak effect on the transverse momentum distributions but they significantly affect the rapidity distribution. Moreover the rapidity distribution of baryons are very sensitive to the values of the parameters *a* and *b* as these are directly related to the chemical freeze-out conditions. Therefore for each hadronic species the values of the parameters are determined from its experimental transverse momentum and the rapidity distributions. This is done in order to understand better the actual thermal conditions prevailing at the time of the freeze-out of any given hadronic specie. For the case of Λ, $\bar{\Lambda}$, Ξ, and $\bar{\Xi}$ the experimental rapidity dN/dy data at √$s_{NN}$ = 200 GeV is available only at the mid rapidity for their ratios i.e. $\bar{\Lambda}$/Λ and $\bar{\Xi}$/Ξ. We have also analyzed and obtained best fits for the experimental transverse momentum and rapidity distribution data of various particles at lower RHIC energy i.e. √$s_{NN}$ = 130 GeV and 62.4 GeV for the 0-5% most central Au+Au collisions. The value of index *n* in our analysis is fixed to be unity and it provides a good fit to the transverse momentum spectra.

The rapidity distributions of protons, antiprotons and $\bar{p}/p$ ratio for the 0-5% most central Au+Au collisions at $\sqrt{s_{NN}}$ = 200 GeV are already reproduced in second chapter. The rapidity distribution of protons were best fitted with the values of the chemical freeze-out parameters a = 22.4 MeV, b = 9.1 MeV and σ = 4.3, while the corresponding thermal/hydrodynamical freeze-out conditions described by T = 162 MeV and $\beta_T^0$ = 0.66 using the minimum $x^2/DoF$ method. For anti-protons these values were found to be T= 163 MeV, $\beta_T^0$ = 0.67, σ = 4.20 and the same values of *a* and *b* as that of the protons. This shows that the protons and anti-protons undergo a near simultaneous freeze-out.

In **Fig.'s 1** and **2** we have shown the transverse momentum distributions for protons and antiprotons at different centralities for Au+Au collisions at $\sqrt{s_{NN}}$ = 200 GeV. We have used the PHENIX Collaboration data [14]. Using our model a best fit to the experimental data is obtained. The theoretical curves obtained by fitting the experimental data are shown by the solid curves. The parameters σ, a and b are determined from the rapidity distributions at $\sqrt{s_{NN}}$ = 200 GeV [12]. These parameters are here then kept fixed and the transverse momentum distribution of protons is studied in different centrality classes. From the best fits using the chi-square minimization method, the behavior of parameters like collective flow velocity and the kinetic temperature with the centrality is studied.

The thermal temperature for both the distributions is found to decrease with the increase in centrality while the collective flow velocity profile increases.

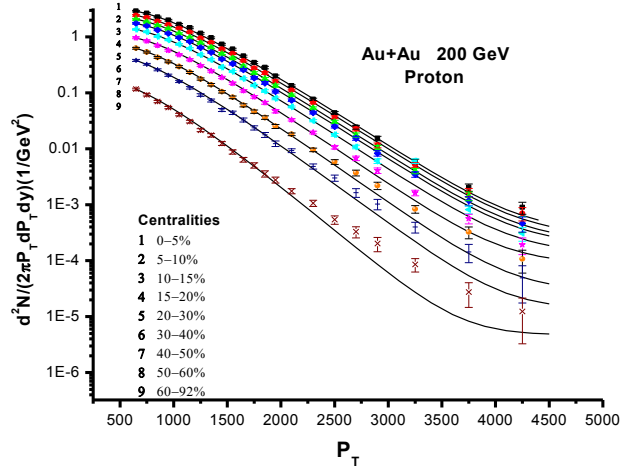

***Fig.1:*** *Transverse momentum distribution of protons at $\sqrt{s_{NN}} = 200$ GeV.*

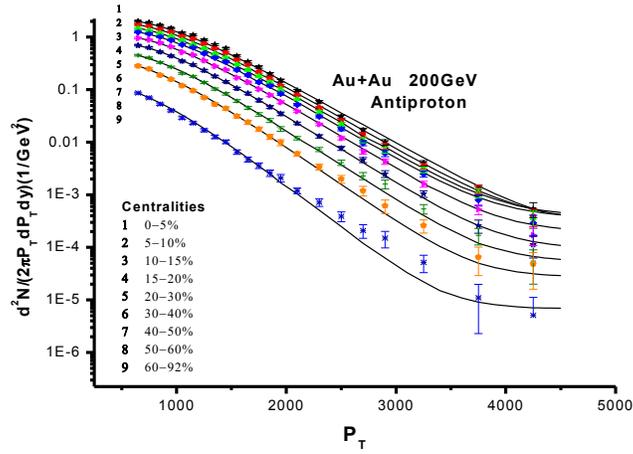

***Fig.2:*** *Transverse momentum distribution of anti-protons $\sqrt{s_{NN}} = 200$ GeV.*

In the above figures the different centrality classes are represented by numbers like 1, 2, 3, … . .The data is taken from PHENIX Collaboration [14]. It is clear that the thermal freeze-out temperature decreases while as the collective flow velocity increases with increasing centrality. The trend of increase/decrease of freeze-out temperature/collective flow velocity parameter with the centrality is shown in the **table 1.1** below:

| Centrality | 0-5 % | 5-10% | 10-15% | 15-0% | 20-30% | 30-40% | 40-50% | 50-60% | 60-92% |
|---|---|---|---|---|---|---|---|---|---|
| T | 162 | 163 | 164 | 164 | 166 | 168 | 169 | 172 | 174 |
| $\beta_T^0$ | 0.66 | 0.66 | 0.65 | 0.65 | 0.63 | 0.60 | 0.56 | 0.51 | 0.40 |

| Centrality | 0-5% | 5-10% | 10-15% | 15-20% | 20-30 | 30-40% | 40-50% | 50-60% | 60-92% |
|---|---|---|---|---|---|---|---|---|---|
| T | 163 | 163 | 164 | 164 | 167 | 168 | 168 | 170 | 171 |
| $\beta_T^0$ | 0.67 | 0.67 | 0.66 | 0.66 | 0.63 | 0.60 | 0.57 | 0.51 | 0.42 |

***Table 1.1:*** *Thermal freeze-out profile for protons (upper table) and antiprotons (lower table)*

It is clear from the tables that the protons and antiprotons have almost simultaneous freeze-out for each different class of centralities and that there is an increase in thermal freeze-out temperature from 163.0 MeV to 171.0 MeV with the decrease in centrality starting from most central 0-5% to the most peripheral 60-92%. The collective flow velocity parameter decreases from 0.67 to 0.42. The increase in the thermal freeze-out temperature may be due to decreasing collective effect as the size of the fireball formed after the collision decreases from central to peripheral collisions. This reduces the multiple collision interaction effect during the evolution of the fireball during which a certain fraction of the thermal energy is converted into directed (hydrodynamic) flow energy. Consequently there is a less drop of temperature in the system before the particles decouple which leads to an early freeze-out of the system. So at this stage the collective flow has not developed significantly in the system and hence shows lower value of $\beta_T^0$.. Next we analyze the transverse momentum ($p_T$) freeze-out spectra of the *strange* particles (mesons as well as singly, doubly and triply strange hyperons and antihyperons) produced in the collisions of Au + Au at $\sqrt{s_{NN}}$ = 200 GeV. In **Fig.'s 3** and **4**, we have shown the transverse momentum distributions of Kaons (K$^+$) and antiKaons (K$^-$) respectively. The already obtained values for the freeze-out parameter values of rapidity distributions Kaons for 0-5 % most central collisions are $\beta_T^0 = 0.56$, T = 161 MeV, σ = 4.10, a = 22.4 MeV and b = 9.1 MeV, while for the anti-Kaons these values turn out to be $\beta_T^0 = 0.58$, T = 161 MeV, σ = 4.36, a = 22.4 MeV and b = 9.1 MeV using the minimum $x^2/DoF$ method.

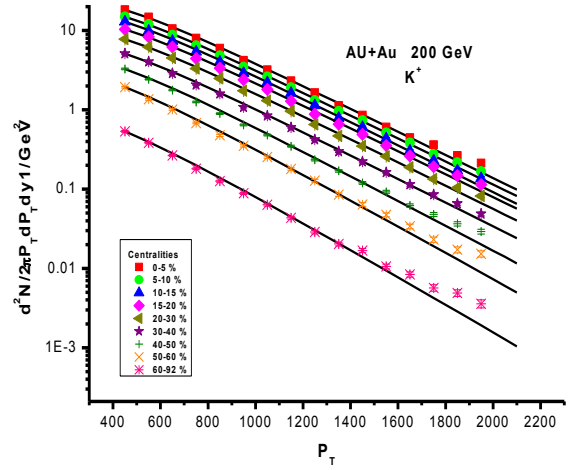

***Fig.3:*** *Transverse momentum distribution of Kaons at $\sqrt{s_{NN}} = 200\ GeV$*

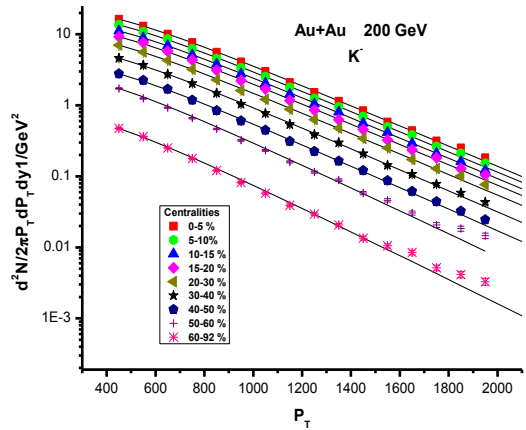

***Fig.4:*** *Transverse momentum distribution of anti-Kaons at $\sqrt{s_{NN}} = 200$ GeV*

However from the study of the transverse momentum distribution of Kaons ($K^+$) at different centralities at $\sqrt{s_{NN}} = 200$ GeV, the values of the thermal freeze-out temperatures for the most central 0-5% and the most peripheral 60-92% collisions are respectively as 161.0 MeV and 175 MeV and the corresponding values of ($\beta_T^0$) are 0.56 and 0.32. For ($K^-$) these values are T = 161.0 MeV and $\beta_T^0 = 0.58$ for the most central 0-5% and T = 173.0 MeV and $\beta_T^0 = 0.37$ for the most peripheral (60-92%) collisions. In **table 1.2** we have shown this trend for the Kaons and antiKaons for varying centralities.

| Centrality | 0-5 % | 5-10% | 10-15% | 15-0% | 20-30% | 30-40% | 40-50% | 50-60% | 60-92% |
|---|---|---|---|---|---|---|---|---|---|
| T | 161 | 164 | 164 | 167 | 167 | 170 | 171 | 175 | 175 |
| $\beta_T^0$ | 0.56 | 0.55 | 0.54 | 0.53 | 0.52 | 0.50 | 0.45 | 0.38 | 0.32 |

| Centrality | 0-5% | 5-10% | 10-15% | 15-20% | 20-30 | 30-40% | 40-50% | 50-60% | 60-92% |
|---|---|---|---|---|---|---|---|---|---|
| T | 161 | 164 | 164 | 165 | 165 | 165 | 166 | 173 | 173 |
| $\beta_T^0$ | 0.58 | 0.57 | 0.57 | 0.56 | 0.55 | 0.53 | 0.51 | 0.41 | 0.37 |

***Table 1.2:*** *Thermal freeze-out profile for Kaons (upper table) and anti-Kaons (lower table)*

The transverse momentum spectra of the singly, doubly and triply strange hyperons are also fitted in the same model. These spectra show a higher freeze-out temperature than other particles (protons and Kaons). The increase in freeze-out temperatures for the strange particles is an indication of their somewhat early freeze-out and hence their spectra exhibit larger thermal temperatures. The transverse momentum distribution for $\Lambda$ and $\overline{\Lambda}$ are shown in the **Fig. 5.** Both $\Lambda$ and $\overline{\Lambda}$ are fitted, for the 0-5% most central collisions, with the same freeze-out conditions, using the collective flow parameter value $\beta_T^0 = 0.60$ and thermal temperature T=167 MeV.

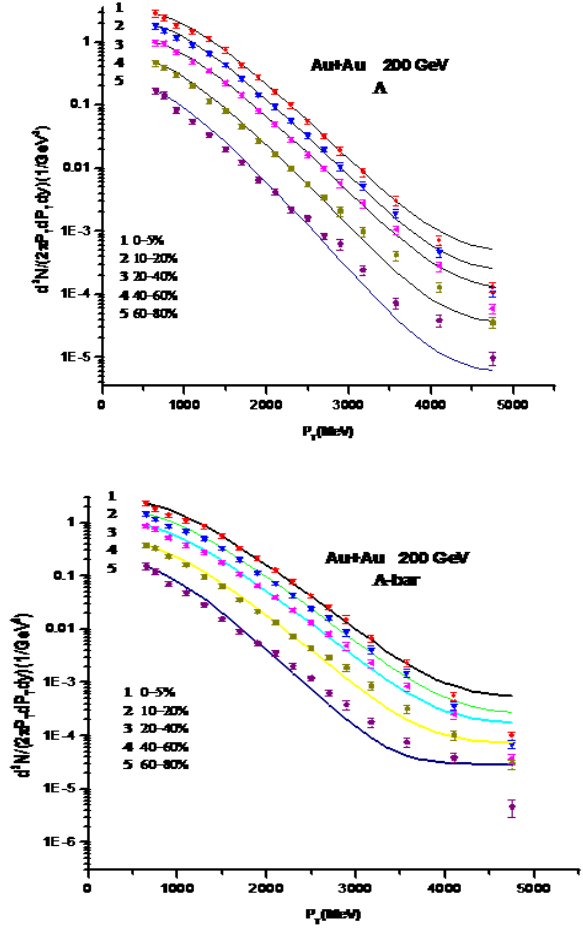

***Fig.5:*** *Transverse momentum distribution of Lambda (upper panel) and anti-Lambda (lower panel)*

The centrality dependence of these parameters for $\Lambda$ and $\overline{\Lambda}$ are exhibited in **table 1.3.**

| Centrality | 0-5 % | 10-20% | 20-40% | 40-60% | 60-80% |
|---|---|---|---|---|---|
| T | 167 | 169 | 171 | 172 | 172 |
| $\beta_T^0$ | 0.60 | 0.59 | 0.59 | 0.53 | 0.45 |

| Centrality | 0-5% | 10-20% | 20-40% | 40-60% | 60-80 |
|---|---|---|---|---|---|
| T | 167 | 169 | 171 | 174 | 176 |
| $\beta_T^0$ | 0.60 | 0.59 | 0.57 | 0.52 | 0.41 |

*Table 1.3: Thermal freeze-out profile for lambda (upper table) and antilambda (lower table)*

The transverse flow parameter $(\beta_T^0)$ values of the lighter particles (i.e. $K^+$ and $K^-$) are found to be smaller than those for the heavier particles (i.e. p, $\bar{p}$, $\Lambda$, $\bar{\Lambda}$ etc;). This suggests that lighter particle's transverse momentum distributions are less influenced by the expansion of the hadronic fluid than the heavier species present in the system.

The transverse momentum distributions for $\Xi^-$ is also well described at different centralities. For the most central collisions the thermal/hydrodynamical freeze-out condition is described by using $\beta_T^0 = 0.60$, T=183 MeV, $\sigma$ = 4.25, a = 22.4 MeV and b = 9.1 MeV. The $\overline{\Xi^=}$ transverse momentum distribution for the most central collisions is fitted by using $\beta_T^0 = 0.61$, T=183 MeV, $\sigma$ = 4.25, a = 22.4 MeV and b = 9.1 MeV. In **Fig. 6**, we have shown the transverse momentum distributions of $\Xi^-$ and $\overline{\Xi^=}$.

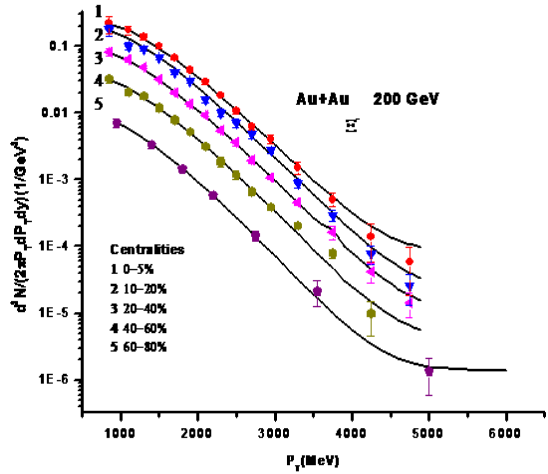

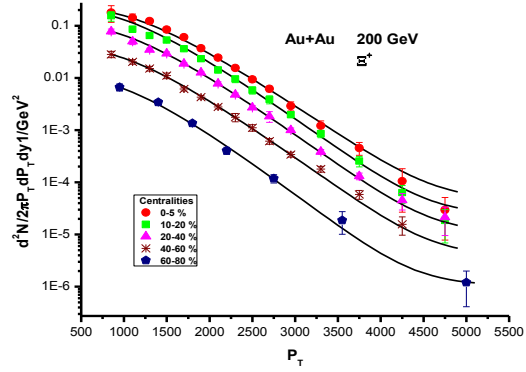

***Fig.6:*** *Transverse momentum distribution of cascade (upper panel) and anti-cascade (lower panel)*

The centrality dependence of these parameters for $\Xi^-$ and $\overline{\Xi^-}$ are shown in **table 1.4.** below.

The freeze-out conditions extracted from the transverse momentum distribution of $(\Omega^- + \Omega^+)$ correspond to $\beta_T^0 = 0.57$ and T=188 MeV.

In **Fig. 7**, we have shown the transverse momentum distributions of $(\Omega^- + \Omega^+)$. The experimental data points are shown by solid circles in all these graphs while the solid curves represent our theoretical results.

| Centrality | 0-5% | 10-20% | 20-40% | 40-60% | 60-80 |
|---|---|---|---|---|---|
| T | 183 | 185 | 190 | 190 | 197 |
| $\beta_T^0$ | 0.60 | 0.55 | 0.55 | 0.53 | 0.47 |

| Centrality | 0-5 % | 10-20% | 20-40% | 40-60% | 60-80% |
|---|---|---|---|---|---|
| T | 183 | 185 | 189 | 190 | 200 |
| $\beta_T^0$ | 0.61 | 0.54 | 0.54 | 0.54 | 0.46 |

***Table 1.4:*** *Thermal freeze-out profile for Cascade (upper table) and anti-Cascade (lower table)*

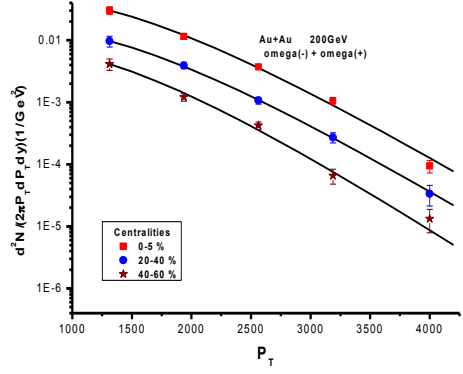

***Fig.7:*** *Transverse momentum distribution of* $(\Omega^- + \Omega^+)$

The centrality dependence of these parameters for $(\Omega^- + \Omega^+)$ are exhibited in **table 1.5.**

| Centrality | 0-5 % | 20-40% | 40-60% |
|---|---|---|---|
| T | 188 | 188 | 190 |
| $\beta_T^0$ | 0.57 | 0.55 | 0.50 |

***Table 1.5:*** *Thermal freeze-out profile for* $(\Omega^- + \Omega^+)$.

Next we study the transverse momentum distribution of identified hadrons at RHIC $\sqrt{s_{NN}}$ = 130GeV center of mass energy. The rapidity distribution of protons, anti-protons and their ratio are shown in the **Fig.'s 8** and **9** below.

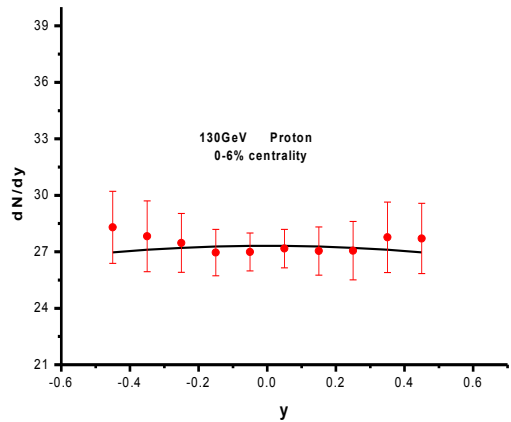

***Fig.8:*** *Rapidity distribution of protons at* $\sqrt{s_{NN}} = 130\ GeV$.

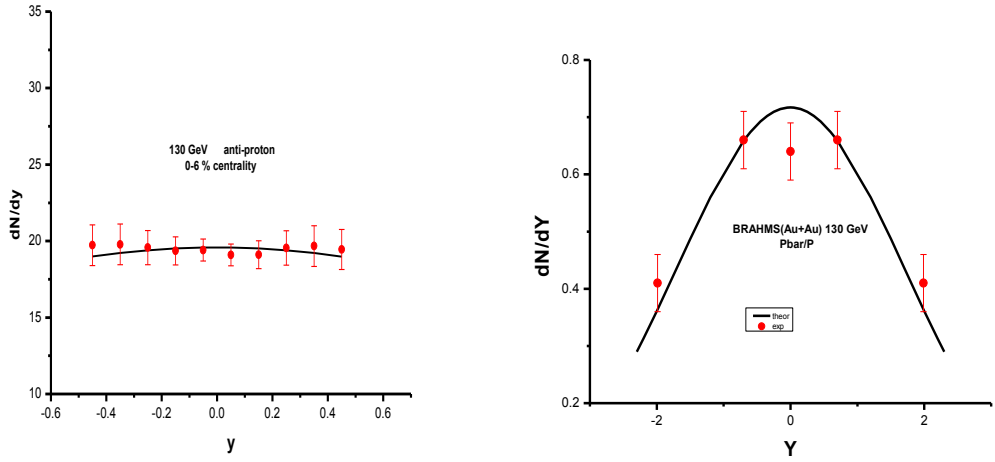

***Fig.9:*** *Rapidity distribution of anti-protons (left) and Pbar/P (Right) at $\sqrt{s_{NN}} = 130$ GeV.*

The value of the parameters a, b, σ, $\beta_T^0$ and T for proton rapidity distribution are respectively as 24.6 MeV, 7.5 MeV, 4.26, 0.55 and 163 MeV. While for antiproton these values are 24.6 MeV, 7.5 MeV, 4.23, 0.55 and 163 MeV. The study of transverse momentum distribution of protons and antiprotons for different centrality are shown in the **Fig.'s 10** and **11**, respectively.

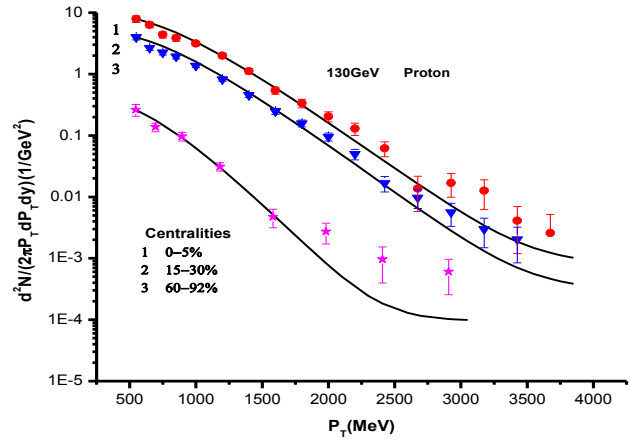

***Fig.10:*** *Transverse momentum distribution of protons at $\sqrt{s_{NN}} = 130$ GeV.*

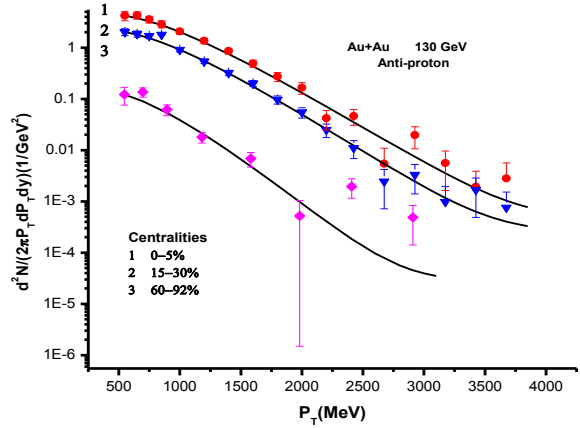

***Fig.11:*** *Transverse momentum distribution of anti-protons at $\sqrt{s_{NN}} = 130$ GeV*

For protons the thermal freeze-out temperature for $\sqrt{s_{NN}} = 130$ GeV slightly increases from 163 GeV to 165 GeV from most central to most peripheral collisions, while there is a decrease in collective flow velocity parameter $\beta_T^0$ from 0.55 to 0.30. This drop in the value of $\beta_T^0$ is significant compared to higher RHIC energy $\sqrt{s_{NN}} = 200$ GeV, where there is a drop in the value of $\beta_T^0$ from 0.66 to 0.40. This shows that the collective flow effects are less developed at the lower energy as compared to $\sqrt{s_{NN}} = 200$ GeV. For antiprotons the freeze-out temperature increases from 163 MeV to 166 MeV and $\beta_T^0$ decreases from 0.55 to 0.31 with the decrease in centrality from most central 0-5% to most peripheral 60-92% collisions. This is depicted in the **table 1.6** below:

| Centrality | 0-5% | 15-30% | 60-92% |
|---|---|---|---|
| T | 163 | 165 | 165 |
| $\beta_T^0$ | 0.55 | 0.50 | 0.30 |

| Centrality | 0-5% | 15-30% | 60-92% |
|---|---|---|---|
| T | 163 | 166 | 166 |
| $\beta_T^0$ | 0.55 | 0.52 | 0.31 |

***Table 1.6:*** *Thermal freeze-out profile for Proton (upper table) and anti-antiproton (lower table) at $\sqrt{s_{NN}} = 130$ GeV.*

The transverse momentum of Kaons and antiKaons are shown below in **Fig. 12.**

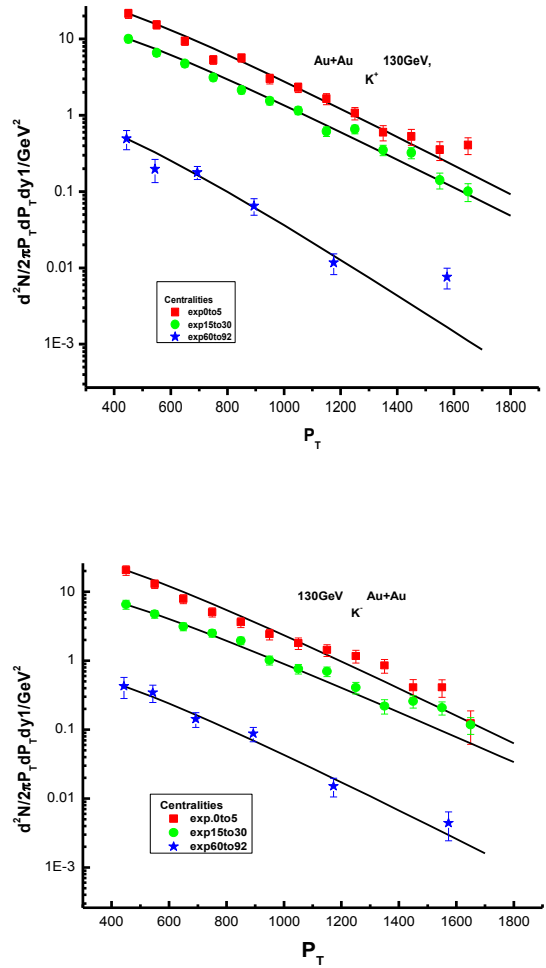

***Fig.12:*** *Transverse momentum distribution of Kaons (upper panel) and antiKaons (lower panel) at $\sqrt{s_{NN}} = 130$ GeV.*

For K$^+$ the thermal freeze-out temperature increases from 164 MeV to 171 MeV while as the collective flow velocity parameter $\beta_T^0$ decreases from 0.50 to 0.30 as we move from most central (0-5%) to the most peripheral (60-92%) collisions. For K$^-$ the thermal freeze-out temperature increases from 161 MeV to 170 MeV and $\beta_T^0$ from 0.50 to 0.32.

The freeze-out thermal temperature for lambda varies from 190 MeV to 207 MeV and the collective flow velocity parameter correspondingly decreases from 0.49 to 0.29 from central to peripheral collisions. For anti-lambda the temperature increases from 190 MeV to 205 MeV and collective flow velocity parameter $\beta_T^0$ decreases from 0.55 to 0.35.

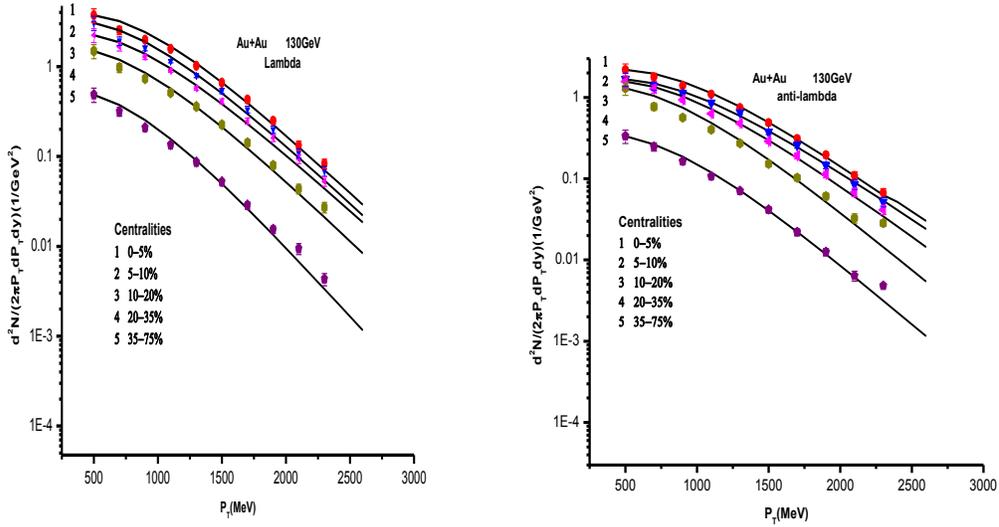

*Fig.13: Transverse momentum distribution of lambdas (left panel) and anti-lambdas (Right panel) at $\sqrt{s_{NN}} = 130$ GeV*

For doubly strange particles (cascade and anti-cascade) the thermal freeze-out temperature is higher than the comparatively lighter particles. For most central 0-10% collisions the $\Xi^-$ and $\Xi^+$ show the freeze-out temperature of 192 MeV and 190.0 MeV while the values of $\beta_T^0$ are 0.48 and 0.47, respectively. This shows that the heavier particles decouple earlier from the system. **Fig. 14** shows their transverse momentum distributions.

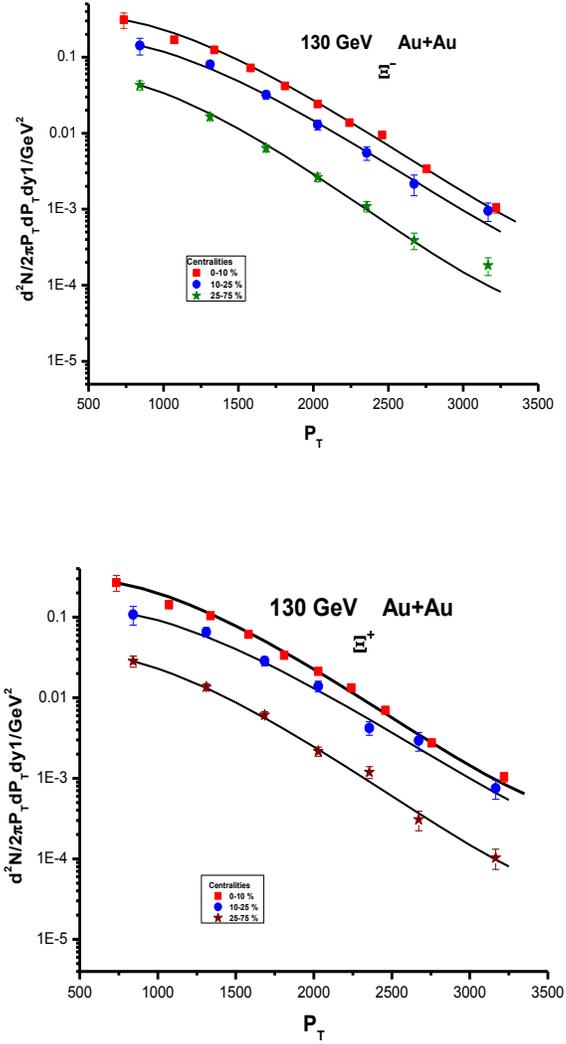

***Fig.14:*** *Transverse momentum distribution of cascade (upper panel) anticascade (lower panel) at $\sqrt{s_{NN}} = 130$ GeV*

The ($\Omega^-$ + $\Omega^+$) transverse momentum spectrum is available only for the central Au + Au collisions. The theoretical fit is shown by the solid line in **Fig. 15.** The thermal freeze-out temperature for this case is 212.0 MeV and the value of the transverse flow parameter $\beta_T^0$ is 0.48.

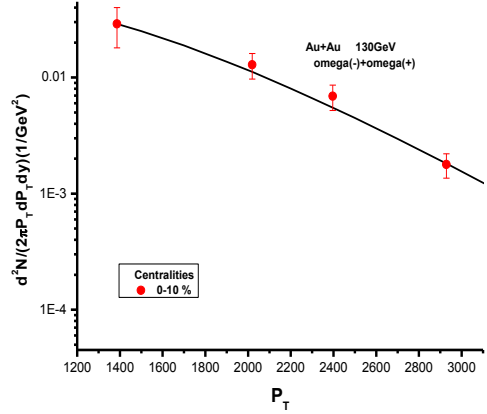

***Fig.15:*** *Transverse momentum distribution of ($\Omega^-$ + $\Omega^+$) at $\sqrt{s_{NN}}$ = 130 GeV.*

We have also studied the distribution of particles at lower RHIC energy i.e. $\sqrt{s_{NN}}$ = 62.4 GeV. The proton (antiproton) rapidity distribution data at this energy is scarce. Further the proton rapidity data does not exhibit the expected trend. Nevertheless, we have still fitted the experimental data using our thermal model. The best fits to the rapidity distributions of proton (antiproton), Kaon (anti-Kaon) and their ratio are shown in the **Fig.'s 16-18**, respectively. The rapidity distribution of proton is fitted with the model parameters *a* = 24.2 MeV, *b* = 25.2 MeV, σ = 4.26, T = 173.0 MeV, $\beta_T^0$=0.54. While the rapidity distribution of anti-proton is fitted with a = 41.0 MeV, b = 6.5 MeV, σ = 2.68, T = 173.0 MeV, $\beta_T^0$=0.55. It is clear from the rapidity distributions of the protons and anti-protons at all energies at $\sqrt{s_{NN}} = 62.4\ GeV$ that the nuclear transparency effect at RHIC increases as we increase the beam energy. This transparency is more exhibited at the highest RHIC energy of $\sqrt{s_{NN}} = 200\ GeV$. The conclusion is drawn from the fact that the value of the parameter "b" increases from 6.5 MeV to 9.1 MeV from the RHIC centre of mass energy of $\sqrt{s_{NN}} = 62.4\ GeV$ to the highest RHIC energy of $\sqrt{s_{NN}} = 200\ GeV$. This means that more the energy of the colliding beams more will be the baryonic matter distributed in the forward rapidity regions of the fireball.

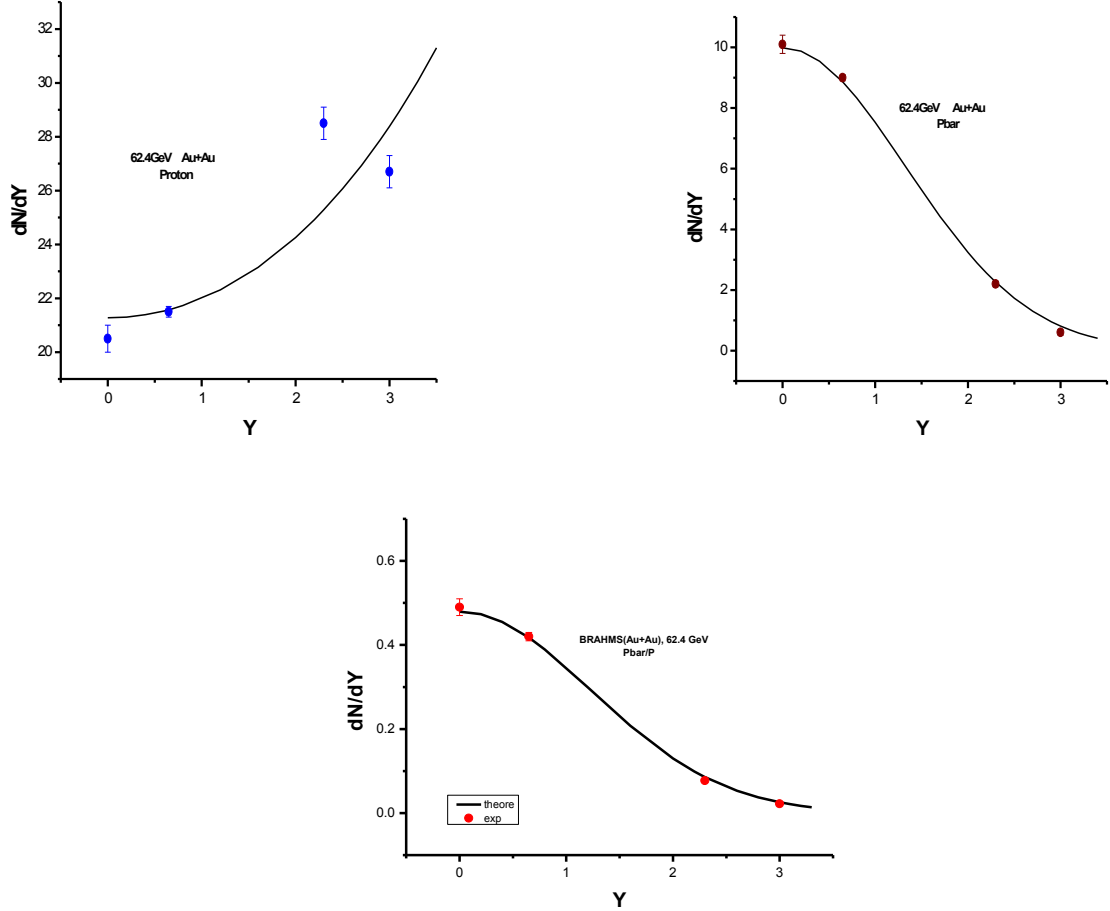

*Fig.16: Rapidity distribution of Proton (left) and anti-Proton (right), lower-panel (Pbar/P) at $\sqrt{s_{NN}} = 62.4\ GeV$*

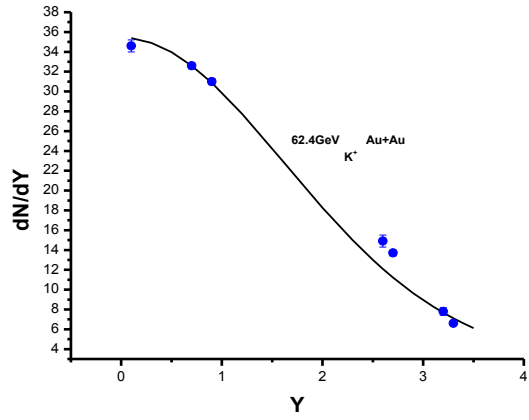

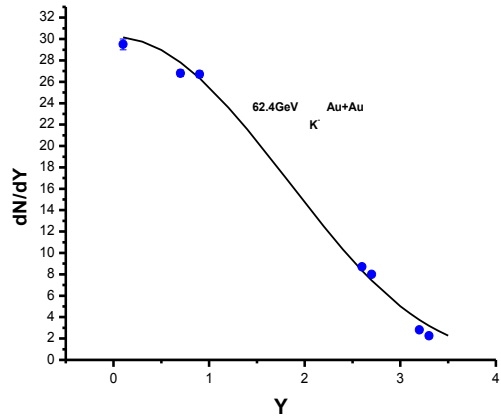

***Fig.17:*** *Rapidity distribution of K⁻(left) and K⁺(right) at $\sqrt{s_{NN}} = 62.4\ GeV$*

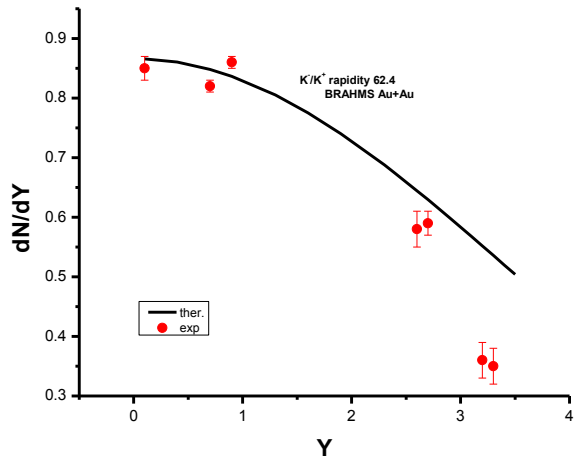

***Fig.18:*** *Rapidity distribution of K⁻/K⁺ at $\sqrt{s_{NN}} = 62.4\ GeV$*

Below we have also shown the transverse momentum distribution of protons, anti-protons, Kaons and anti-Kaons at the lower RHIC energy i.e. $\sqrt{s_{NN}}$ = 62.4 GeV. In **Fig.'s 19** and **20** the transverse momentum spectra of protons and anti-protons.

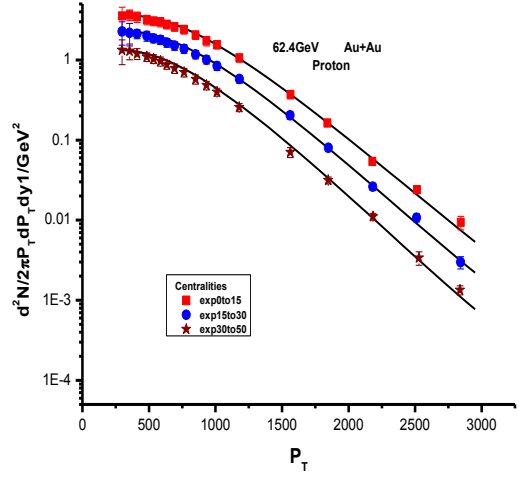

***Fig.19:*** *Transverse momentum distribution of Protons at* $\sqrt{s_{NN}} = 62.4\ GeV$

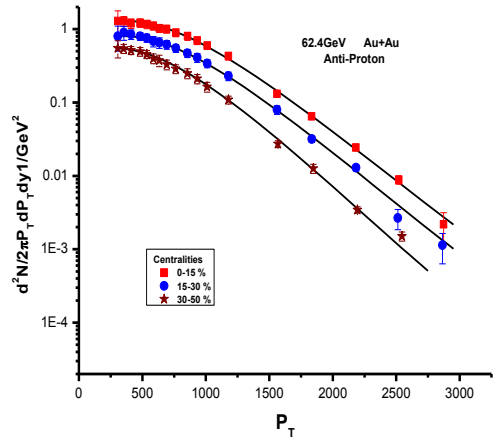

***Fig.20:*** *Transverse momentum distribution of antiprotons at* $\sqrt{s_{NN}} = 62.4\ GeV$

The profile of the freeze-out parameters obtained from the transverse momentum distribution of protons and anti-protons at different centralities with our model calculation at RHIC $\sqrt{s_{NN}} = 62.4\ GeV$ energy are shown in the **table 1.7** below:

| Centrality | 0-15% | 15-30% | 30-50% |
|---|---|---|---|
| T | 173 | 174 | 176 |
| $\beta_T^0$ | 0.54 | 0.50 | 0.44 |

| Centrality | 0-15% | 15-30% | 30-50% |
|---|---|---|---|
| T | 173 | 174 | 175 |
| $\beta_T^0$ | 0.55 | 0.52 | 0.42 |

*Table 1.7:* *Thermal freeze-out profile for Proton (upper table) and anti-Proton (lower table) at different centralities at $\sqrt{s_{NN}} = 62.4\ GeV$.*

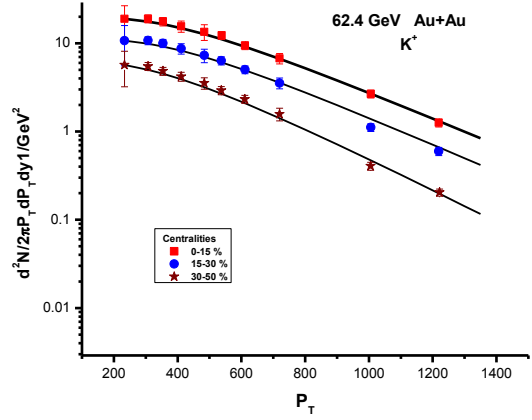

*Fig.21*: *Transverse momentum distribution of Kaons at $\sqrt{s_{NN}} = 62.4\ GeV$*

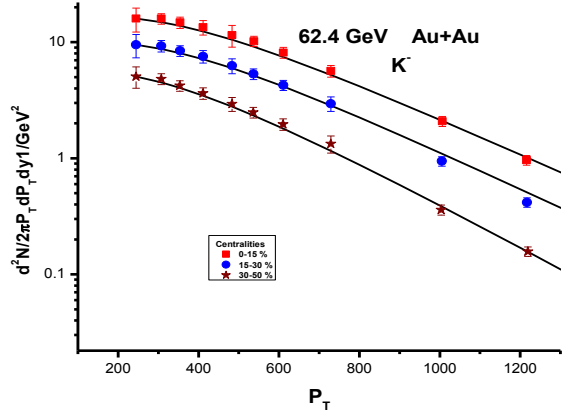

***Fig.22:*** *Transverse momentum distribution of antiKaons at $\sqrt{s_{NN}} = 62.4\ GeV$*

The profile of the freeze-out parameters obtained from the transverse momentum distribution of kaons and anti-Kaons at different centralities with our model calculation at RHIC $\sqrt{s_{NN}} = 62.4\ GeV$ energy are shown in the **table 1.8** below:

| Centrality | 0-15 % | 15-30% | 30-50% |
|---|---|---|---|
| T | 169 | 173 | 175 |
| $\beta_T^0$ | 0.49 | 0.48 | 0.28 |

| Centrality | 0-15% | 15-30% | 30-50% |
|---|---|---|---|
| T | 168 | 168 | 170 |
| $\beta_T^0$ | 0.49 | 0.44 | 0.27 |

***Table 1.8:*** *Thermal freeze-out profile for Kaon (upper table) and anti-Kaon (lower table) at different centralities at $\sqrt{s_{NN}} = 62.4\ GeV$.*

The transverse momentum distribution of Λ, anti-Λ, Ξ⁻ and Ξ⁺ are shown in the **Fig's 23 to 26**. The thermal freeze-out temperature for the lambda particle increases from 196 MeV to 200 MeV from most central (0-5 %) to most peripheral (60-80 %) respectively while as the collective flow velocity parameter decreases from 0.45 to 0.32. Almost the similar freeze-out takes place for the most centrality for anti-lambda and for the most peripheral collision the freeze-out condition are T=202 MeV and $\beta_T^0 = 0.22$.

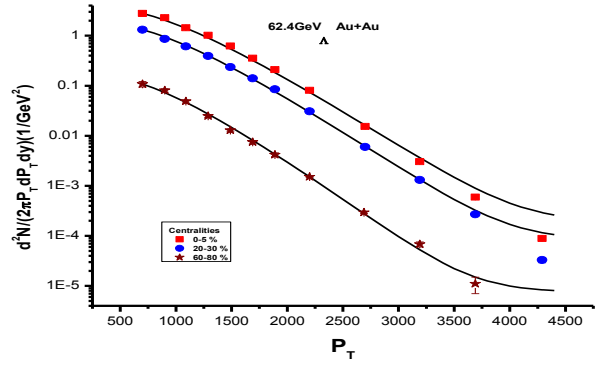

***Fig.23:*** *Transverse momentum distribution of Λ at $\sqrt{s_{NN}} = 62.4\ GeV$.*

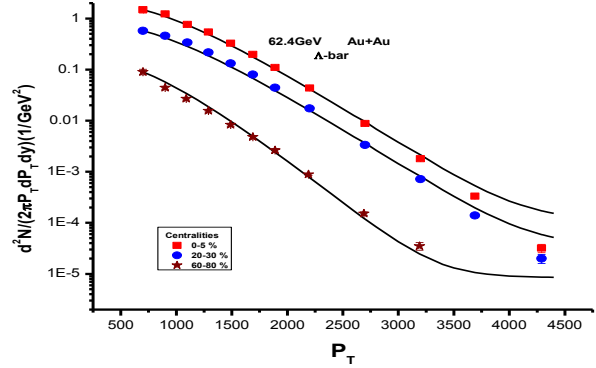

***Fig.24:*** *Transverse momentum distribution of anti-Λ at $\sqrt{s_{NN}} = 62.4\ GeV$.*

The transverse momentum distribution of cascade particle and anti-cascade particle at different centralities are shown in **Fig.'s 25** and **26** respectively.

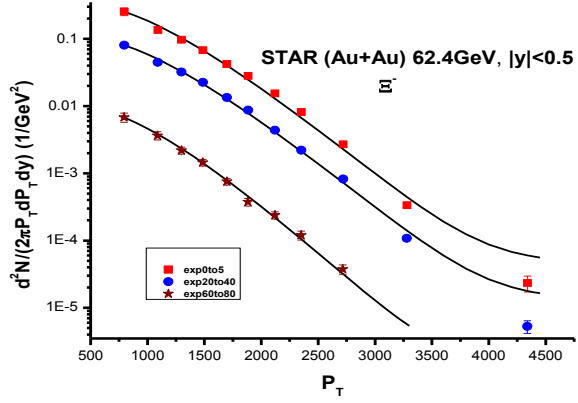

***Fig.25:*** *Transverse momentum distribution of* $\Xi^-$ *at* $\sqrt{s_{NN}} = 62.4\ GeV$.

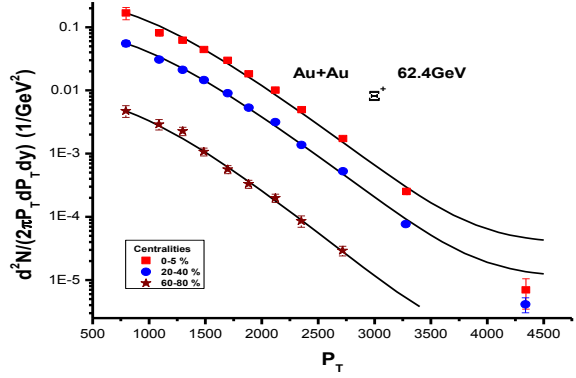

***Fig.26:*** *Transverse momentum distribution of* $\Xi^+$ *at* $\sqrt{s_{NN}} = 62.4\ GeV$.

The profile of freeze-out conditions for the **Fig.'s 25** and **26** are shown in **table 1.9** below:

| Centrality | 0-5 % | 20-40% | 60-80% |
|---|---|---|---|
| T | 205 | 208 | 209 |
| $\beta_T^0$ | 0.37 | 0.36 | 0.25 |

| Centrality | 0-5% | 20-40% | 60-80% |
|---|---|---|---|
| T | 204 | 208 | 208 |
| $\beta_T^0$ | 0.37 | 0.35 | 0.29 |

*Table 1.9: Thermal freeze-out profile for Cascade (upper table) and anti-Cascade (lower table) for different centralities at $\sqrt{s_{NN}} = 62.4\ GeV$.*

The transverse momentum distribution of heavy baryons $\Omega^-$ and $\Omega^+$ are also shown in the **Fig.'s 27** and **28** respectively. The kinetic freeze-out temperature and collective velocity flow parameter for $\Omega^-$ for the most central collision (0-20 % centrality) are respectively 208 MeV and 0.33. While for the most peripheral collision (40-60%), these values turn out to be respectively 212 MeV and 0.30. This shows the same trend of increasing temperature and decreasing collective velocity flow parameter with the decrease in centrality even at lower RHIC energy in accordance with the heavy baryon freeze-out discussed at higher RHIC energies. The freeze-out condition for the most central collision for $\Omega^+$ at the same energy are respectively T=208 MeV and $\beta_T^0$=0.35. The thermal freeze-out temperature at this energy also is higher than the thermal freeze-out temperature of doubly strange particles.

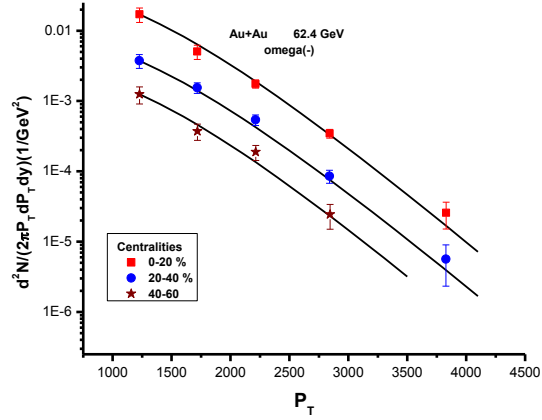

***Fig.27:*** *Transverse momentum distribution of $\Omega^-$ at $\sqrt{s_{NN}} = 62.4\ GeV$.*

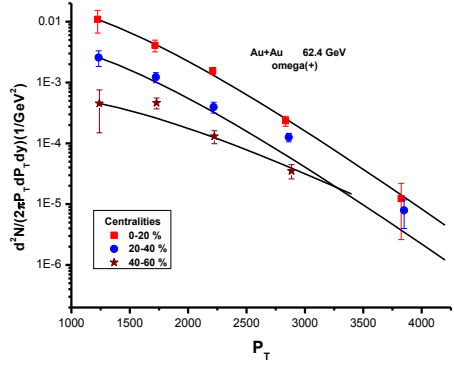

***Fig.28:*** *Transverse momentum distribution of $\Omega^+$ at $\sqrt{s_{NN}} = 62.4\ GeV$.*

## 1.3  Summary and Conclusion

We have successfully described the transverse momentum and rapidity spectra of various hadrons produced in Au + Au collisions at various RHIC energies of $\sqrt{s_{NN}} = 200, 130$ and $62.4$ GeV. Our model incorporates longitudinal as well as transverse boosts developed during the evolution of the fireball formed in the Au + Au relativistic collisions. Our analysis showed that the spectra of almost all hadronic species can be well described when accounting the collective flow of particles. It is concluded that the heavier or strange hadrons decouple earlier from the system than the non-strange ones. Dependence of the freeze-out parameters on the degree of centrality of Au+Au collisions at various RHIC energies is also described and it is found that the kinetic temperature increases and collective flow effects decreases with decreasing centrality at all energies studied. The nuclear transparency effect is also studied and it is clear from our model that the nuclear matter becomes more transparent at the highest RHIC energy compared to lower RHIC energies.

## 1.4 References

[1]  Norbert Hermann, Johannes P. Wessels and Thomas Wienold

 *Annu.Rev.Part.Sci.*1999.49:581-632

[2]  ALICE COLLABORATION, *Phys. Rev. C* 88, 044909(2013)

[3]  Peter Braun-Munzinger, Krzysztof Redlich, Johanna Stachel, *arxiv:*nucl-th

 /0304013v1(2013)

[4]  V. P. Kondrat'ev, G. A. Feofilov, 2011, published in Fizika Elementarnykh Chastits I

 atomnogo yadra, 2011, Vol.42, No.06

 [5]  V.K. Magas, L.P. Csernai, D.D. Strottman, *Phys. Rev. C* **64** (2001) 014901;

  *Nucl. Phys*. **A712** (2002) 167; E. Kornas *et al*. NA 49 Collaboration, *Eur. Phys. J.*

 **C49** (2007) 293

[6]  Saeed Uddin, N. Akhtar and M. Ali, *Int. J. Mod. Phys*. **21**, 1472 (2006)

[7]  Saeed Uddin, Jan Shabir Ahmad, Waseem Bashir and Riyaz Ahmad Bhat.

 *J. Phys.* **G 39**, 015012 (2012)

[8]  Adler C et al STAR Collaboration *Phys. Rev. Lett. 87*, 262302, (2001).

[9]  Becattini F *et al* *Phys. Rev. C* 64, 024901, (2001).

[10]   Braun-Munzinger P, Magestro D, Redlich K and Stachel J 2001 *Phys.Lett. B* 518 41

[11]   Florkowski W, Broniowski W and Michalec M 2002 *Acta Phys.Pol.B* 33 761

[12]   Saeed Uddin, Riyaz Ahmed Bhat, Inam-ul Bashir, Waseem Bashir, Jan Shabir Ahmad *arxiv:* 1401.0324[hep]

[13]   F.Becattini, J. Cleymans and J. Strmpfer, arXiv: 0709.2599 v1[hep-ph]

[14]   PHENIX Collaboration, S.S.Adler, *et al* 10.1103/*Phys. Rev. C*.69.034909.